\documentclass[12pt]{iopart}

\usepackage{iopams}
\usepackage{bm}
\usepackage[colorlinks=true, urlcolor=blue, linkcolor=blue, citecolor=blue, pdftex]{hyperref}
\usepackage{float}
\usepackage[dvipsnames]{xcolor}
\usepackage{comment}
\usepackage[utf8]{inputenc}
\usepackage{braket}
\usepackage{graphicx}
\usepackage{amsmath}

\renewcommand{\Braket}[2]{\left\langle #1 \middle| #2 \right\rangle}
\newcommand{\Av}[1]{\left \langle #1 \right \rangle}

\newcommand{\VO}[1]{\textcolor{purple}{#1}}

\begin{document}

\title{A kinetically constrained model exhibiting non-linear diffusion and jamming}
\author{Abhishek Raj,    Vadim Oganesyan}
\address{Physics Program and Initiative for the Theoretical Sciences, The Graduate Center, CUNY, New York, New York 10016, USA}
 \address{Department of Physics and Astronomy, College of Staten Island, CUNY, Staten Island, New York 10314, USA
 }
\author{Antonello Scardicchio}
\address{The Abdus Salam ICTP, Strada Costiera 11, 34151, Trieste (Italy)\\ 
INFN, Sezione di Trieste, Via Valerio 2, 34128, Trieste (Italy).}

\vspace{10pt}

\begin{abstract}
We present a classical kinetically constrained model of interacting particles on a triangular ladder, which displays diffusion and jamming and can be treated by means of a classical-quantum mapping. Interpreted as a theory of interacting fermions, the diffusion coefficient is the inverse of the effective mass of the quasiparticles which can be computed using mean-field theory. At a critical density $\rho=2/3$, the model undergoes a dynamical phase transition in which exponentially many configurations become jammed while others remain diffusive. The model can be generalized to two dimensions.                                                                                 
\end{abstract}

\section{Introduction}

In recent years, we have witnessed a shift of attention in the field of statistical physics from the study of thermodynamic, equilibrium properties, which have been the core of the subject since its inception, to the characterization of the approach to equilibrium, or lack thereof. This is due to theoretical and experimental advances, both in quantum and classical systems. In the first realm, the current degree of control on cold atomic arrays \cite{gross2017quantum}, superconducting qubits arrays \cite{kjaergaard2020superconducting,georgescu2014quantum}, and the larger availability of supercomputing resources has spurred a renewed interest in the dynamics of many-body quantum systems and in particular in their approach to equilibrium. The eigenstate thermalization hypothesis \cite{deutsch2018eigenstate,srednicki1994chaos,d2016quantum}, and its counterpart for strongly disordered systems, many-body localization \cite{basko2006metal,oganesyan2007localization,abanin2019colloquium, nandkishore2015many,sierant2024many} are two prominent examples of this work. In the classical world, the line of work which started with the study of equilibrium properties of spin glasses \cite{edwards1975theory,mezard1987spin} turned out to apply well beyond the original scope \cite{kirkpatrick1994critical}, and evolved into a theory of complexity \cite{parisi2023nobel}. The mean-field theory, which comprises replica symmetry breaking, has been successfully applied to computational complexity \cite{mezard2002analytic, mezard2002random}, structural glasses \cite{mezard2012glasses}, and many other examples \cite{parisi2010mean}. In this paper we turn our attention to a classical random process model which has slow dynamics and jamming, and whose solution is obtained by mapping to a model of interacting fermions, thereby connecting the two lines of work.

\begin{figure}
    \centering
\includegraphics[width=0.6\columnwidth]{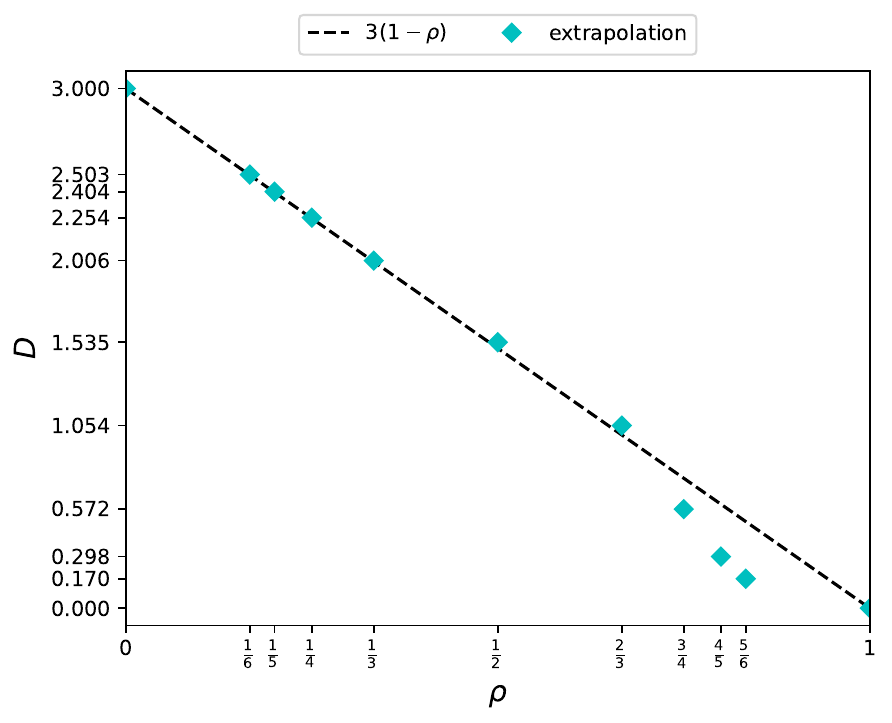}
    \caption{
Comparison of mean-field solution (dashed straight-line) against extrapolated numerical results (green diamonds)  -- see Figs. \ref{fig:low_density_1bn} and \ref{fig:high_density};  
Notice that the numerical data follow the mean-field prediction of Eq.(\ref{eq:DMF}) quite closely until the jamming point $\rho=2/3$.  The disagreement at high density may be due to progressively less accurate extrapolation of numerics.
}
\label{fig:Particle_sector}
\end{figure}

To achieve this, we consider a model of interacting random walkers with simple constraints, or a {\it kynetically constrained process} (KCP). KCP's are a widely used tool in statistical mechanics to describe the approach to equilibrium, the emergence of non-equilibrium states, and glassy dynamics \cite{DERRIDA199865,ritort2003,bertini2007stochastic}. Among other aspects, they are useful to study the emergence, in a controlled setting, of the hydrodynamic behavior and the statistics of large deviations from equilibrium. Such family of models include those introduced and treated by Katz, Lebowitz and Spohn \cite{katz1984nonequilibrium} and many others \cite{eyink1990hydrodynamics,spohn2012large}. The typical strategy to find the diffusion coefficient in these models is to use the Green-Kubo formula and thermodynamics to compute it, or if this is not possible, bound  \cite{krapivsky2012fluctuations,krapivsky2013dynamics}. This is not the route we will follow in this paper.

In this paper, we introduce and study a simple exclusion process in which particles can hop to their left or right conditioned to the fact that {\it both} the left and right side is empty. The model is easily defined on a triangular {\it ladder}, which will allow, in further work, to define it on a two-dimensional, triangular {\it lattice}. 

We employ a classical-to-quantum mapping to turn our KCP into a frustration-free\cite{han2024models} spin Hamiltonian whose groundstates may be obtained exactly\cite{han2024models}. To demonstrate the existence of a diffusive behavior in the original model we study the dispersion of elementary excitations of the quantum problem -- the diffusion coefficient is proportional to the inverse mass of the quasiparticle. By construction, the diffusion coefficient is expected to decrease monotonically with increasing particle density $\rho$.  

We find the diffusion coefficient with a mean-field treatment of the fermionic Hamiltonian and compare against exact numerical results on finite chains, see Fig. \ref{fig:Particle_sector}. The mean-field solution $D=3(1-\rho)$, where $\rho$ is the density of particles, agrees surprisingly well with the numerics in a very large range of densities, see Fig.\ref{fig:Particle_sector}, which makes us suspect it is an exact solution. 

This is more surprising if we consider that the model exhibits the presence of jammed configurations, which appear at a critical density $\rho=2/3$, therefore showing an interesting connection with the study of jamming in granular materials \cite{donev2004jamming,parisi2010mean,behringer2018physics}.

\section{Preliminaries -- model, mapping, definitions}
\subsection{The model and a classical-quantum mapping}
\label{sec:model}
\begin{figure}
    \centering
    \includegraphics[width=0.5\linewidth]{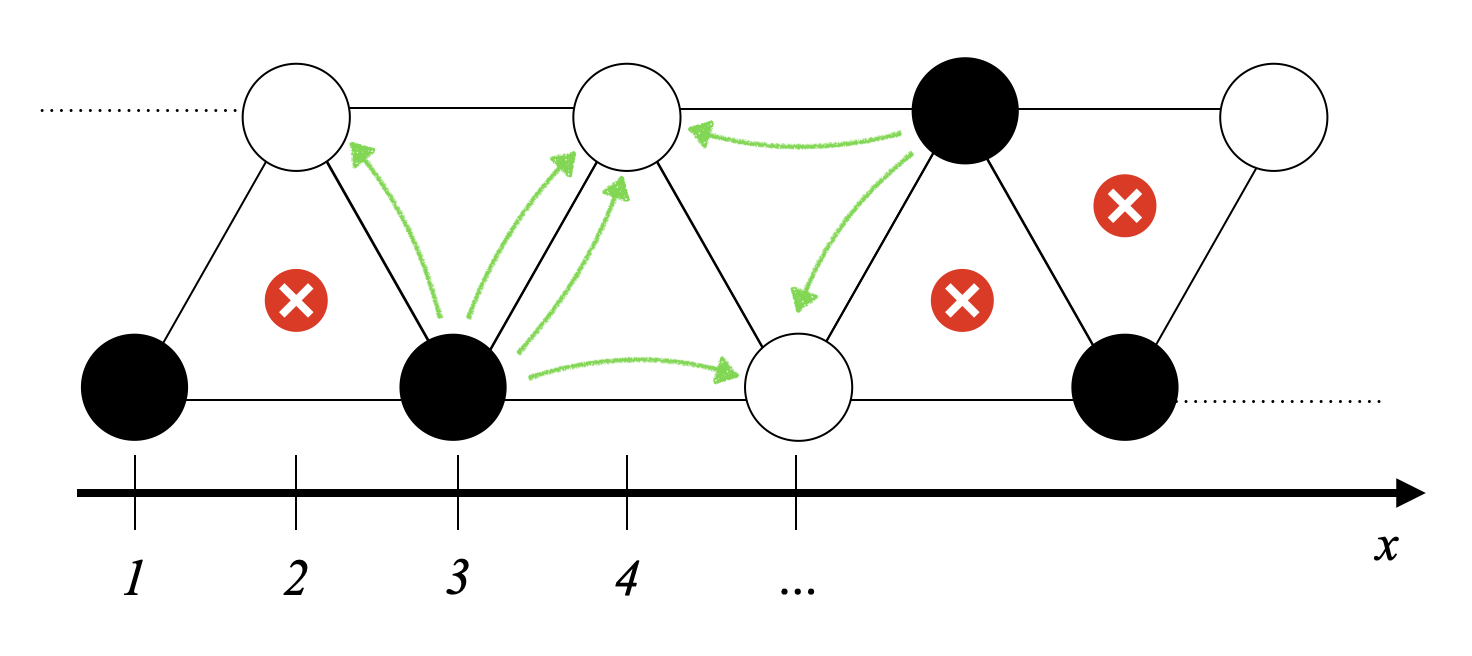}
    \caption{The rules for hopping. In active triangles, where only one particle out of 3 is present, it can hop with the same probability along the green arrows. Where two or more particles are present in a triangle (red cross) such triangle is inactive. On the lower axis, the numeration convention we use.}
    \label{fig:hopping}
\end{figure}

The random process studied describes random updates of 3 particles configurations where the only mobile configurations are those in which a particle is free on both sides. The rationale behind this choice\cite{raj2024diffusion} is to have a simple explicit model of random walkers with strongly dependent diffusion constant that produces strong nonlinear interactions among hydrodynamic modes.
%avoid any particle-hole symmetry, in such a way to guarantee a monotonically decreasing diffusion coefficient. 
The rule is sketched in Figure \ref{fig:hopping}. It is also straightforward to implement in large scale simulations -- see e.g. Ref.\cite{raj2024diffusion}. This work, by contrast, focuses on exact noise-averaged formulation of the dynamics as we explain now. 
%\\
%\VO{VO:Bring the first paragraph from the NuMX section here -- to contrast noisy vs. noise free simulation options?}

In order to write the rate matrix we choose the basis $000,001,...,111$, where $0$ is an empty site and $1$ a particle. On these 3 neighboring lattice sites, the matrix of the transition rates is
\begin{equation}
    h_{i,i+1,i+2}=\left(
\begin{array}{cccccccc}
 0 & 0 & 0 & 0 & 0 & 0 & 0 & 0 \\
 0 & -1 & \frac{1}{2} & 0 & \frac{1}{2} & 0 & 0 & 0 \\
 0 & \frac{1}{2} & -1 & 0 & \frac{1}{2} & 0 & 0 & 0 \\
 0 & 0 & 0 & 0 & 0 & 0 & 0 & 0 \\
 0 & \frac{1}{2} & \frac{1}{2} & 0 & -1 & 0 & 0 & 0 \\
 0 & 0 & 0 & 0 & 0 & 0 & 0 & 0 \\
 0 & 0 & 0 & 0 & 0 & 0 & 0 & 0 \\
 0 & 0 & 0 & 0 & 0 & 0 & 0 & 0 \\
\end{array}
\right)
\end{equation}
The total rate matrix is
\begin{equation}
    H=p \sum_{i=1}^L h_{i,i+1,i+2}
    \label{eq:H}
\end{equation}
where we use periodic boundary conditions, and it is a positive matrix, whose steady state (later, quantum ground state energy) corresponds to the zero eigenvalues, $E_0=0$. Master equation 
%$H$ rules 
governs the evolution of probabilities of particle configurations $\sigma=\{\sigma_1,...,\sigma_L\}\in \{0,1\}^L$ 
\begin{equation}
    \dot{P}(\sigma,t)=-\sum_{\sigma,\sigma'}H_{\sigma,\sigma'} P(\sigma',t).
    \label{eq:master}
\end{equation}
Time-dependent expectation values of observables $O$ can be computed as
\begin{eqnarray}
    \Av{O(t)}=\sum_{\sigma}O(\sigma) P(\sigma,t)&=&\sum_{\sigma,\sigma'}O(\sigma) \left(e^{-Ht}\right)_{\sigma,\sigma'}P(\sigma',0),\label{eq:evolA}\\
    &=&\sum_{a}e^{-E_{a} t}\sum_{\sigma,\sigma'}O(\sigma)\Braket{\sigma}{E_{a}}\Braket{E_{a}}{P(0)}\label{eq:evolB},
\end{eqnarray}
where $\ket{E_a}$ are the eigenvectors of $H$. Note: while Eq.(\ref{eq:evolA}) is a general solution of Eq.(\ref{eq:master}), Eq.(\ref{eq:evolB}) only follows for problems obeying detailed balance, {\it i.e.} the rate matrix is symmetric and therefore admits left-right-symmetric eigen-decomposition.  Finally, we note the value of the diffusion constant of a single random walker on this lattice in discrete time is $D_0=(p/2) (2\cdot (2a)^2+4\cdot a^2)/(2\Delta t)$, as each site belongs to 3 triangles with 6 possible hops. In what follows we set $p=a=\Delta t=1$, for simplicity and $D_0=3$.

\subsection{Classical-to-quantum mapping}
In what follows we will treat $H$ in Eq. \ref{eq:H} as a quantum many-body Hamiltonian. To make the mapping complete we need to reinterpret Eq. \ref{eq:master} as governing the evolution of $\mathcal{L}^2-$ normalized states obtained by taking the square root of the (non-negative) probability distribution (which by itself is $\mathcal{L}^1-$ normalized).  The central benefit of doing this is to organize calculations using well established language and formalism of many-body theory, including use of exact properties of eigenstates and eigenvalues, Jordan-Wigner transform and approximation schemes (specifically Hartree approximation used below).  Additionally, this formulation also streamlines exact numerical analysis finite-size lattices -- see below and also Ref.\cite{raj2024diffusion}.  Others, e.g. Ref. \cite{DieterJ_TEBDStochastic}, have exploited this mapping to take advantage of matrix-product based tools for computing temporal evolution in stochastic problems.

%The Hamiltonian in Eq. \ref{eq:H} commutes with the total number of particles $N$ and the translation operator $T$ which rotates the particle configuration by a lattice spacing $T\ket{\sigma_1,...,\sigma_L}=\ket{\sigma_2,...,\sigma_L,\sigma_1}$. All eigenstates can then by labeled by $N,Q$ and the set of linear superpositions over configurations with $N$ particles is $\mathbb{B}_{N,L}=\{\sigma\in\{0,1\}^L|\sum_i\sigma_i=N\}$. The equilibrium (infinite temperature) distribution corresponds to the ground state in the $Q=0$ sector, $T\Ket{E_{0,Q=0}}=\Ket{E_{0,Q=0}}$ and $E_{0,Q=0}=0$.  The ground state, owning to the double-stochastic condition $\sum_{\sigma}H_{\sigma',\sigma}=0$, is the uniform superposition of all the particle configurations
%\begin{equation}
%    \Ket{E_{0,N}}=\frac{1}{Z^{1/2}}\sum_{\sigma\in \mathbb{B}_{N,L}}\Ket{\sigma},
%\end{equation}
%where $Z=\binom{L}{N}$. 

The Hamiltonian in Eq. \ref{eq:H} commutes with the total number of particles $N$ and the translation operator $T$ which rotates the particle configuration by a lattice spacing $T\ket{\sigma_1,...,\sigma_L}=\ket{\sigma_2,...,\sigma_L,\sigma_1}$. All eigenstates can then by labeled by $N$, $Q$ (the lattice momentum) and eigenvalues of $H$ (``energy").
%--  and the set of linear superpositions over configurations with $N$ particles is $\mathbb{B}_{N,L}=\{\sigma\in\{0,1\}^L|\sum_i\sigma_i=N\}$. 

The equilibrium (infinite temperature) distribution corresponds to the ground state in the $Q=0$ sector, $T\Ket{E_{0,N,Q=0}}=\Ket{E_{0,N, Q=0}}$ and since $E_{0,N,Q=0}=0$ we have that the ground state in the $N$ particle sectors belongs to the $Q=0$ sector. We define this ground state $\ket{E_{0,N}}=\ket{E_{0,N,Q=0}}$, which owning to the double-stochastic condition $\sum_{\sigma}H_{\sigma',\sigma}=0$, is the uniform superposition of all the $N$-particle configurations
\begin{equation}
    \Ket{E_{0,N}}=\frac{1}{Z^{1/2}}\sum_{\sigma\in \mathbb{B}_{N,L}}\Ket{\sigma},
\end{equation}
where $\mathbb{B}_{N,L}=\{\sigma\in\{0,1\}^L|\sum_i\sigma_i=N\}$ and $Z=\binom{L}{N}$.

Any state $\ket{P(t)}$ has the same overlap with $\ket{E_{0,N}}$ by normalization:
\begin{equation}
    \Braket{E_{0,N}}{P(t)}=\frac{1}{Z^{1/2}}\sum_{\sigma\in\mathbb{B}_{N,L}}P(\sigma,t)=\frac{1}{Z^{1/2}},
\end{equation}
so at asymptotically large times, we have, irrespective of $P(0)$,
\begin{equation}
    \Av{O(t)}\simeq \frac{1}{Z}\sum_\sigma O(\sigma)+c_1 e^{-E_1 t}+...\ ,
\end{equation}
where $c_1\propto \Braket{E_1}{P(0)}$ and $E_1>0$ refers to the lowest excited state. This Hamiltonian commutes with both particle number $N$ and lattice momentum $Q$ so we can focus on individual $N,Q$ sectors, which is especially useful in finite lattices (see below).

%\textcolor{red}{[AS: The necessity to define $E_{0,N}$ before, led to the merging of sections 2.1 and 2.2. \\VO: I see...I think there is value in a clear separation between classical and quantum, at the very least to avoid any further confusion, so I will go back to the current eqs 3 and 4, and explain the expansion of the evolution operator into modes (zero and otherwise) to make it clear that it exists without quantum mapping]}

%

Lastly, it is important to keep in mind that while $H$ commutes with $N,T$ (and, trivially, itself), under imaginary time evolution that interests us, only the $N$ is conserved. This is due to the fact that (for any $N$) only the $Q=0$ ground state %(with different particle numbers $N$) 
has zero energy thus representing stochastic steady state.  Ground states in finite $Q$ sectors will see their amplitude in the vector $P$ decay like $e^{-E_{0,N,Q} t}$. 
%I one does observe that $E_{0,Q} \simeq D Q^2$ in the low $Q$ limit, this defines a diffusion constant $D$. 
We now turn to precise definition and method of extracting $D$ from these decays. %and associated nuances, both general and in this model.

%and is a trivial representation of the translation group ({\it i.e.} $T\Ket{E_0}=\Ket{E_{0,Q=0}})$. The ground state, due to the condition $\sum_{\sigma}H_{\sigma',\sigma}=0$, is the uniform superposition of all the particle configurations:
%\begin{equation}
%    \Ket{E_{0,N}}=\frac{1}{Z^{1/2}}\sum_{\sigma\in %\mathbb{B}_{N,L}}\Ket{\sigma},
%\end{equation}
%where $Z=\binom{L}{N}$.

\subsection{D-definitions}

For completeness we now recall the precise and completely general definition of the diffusion constant\cite{Chaikin_Lubensky_1995} to elucidate the importance of the unique correct order of limits involved. We then highlight other dynamical processes that take place in interacting diffusive systems (including in this model) such as long-time tails (LTT) and diffusion cascade\cite{heavyluca,raj2024diffusion}.

The conventional definition of diffusion constant of the test (Brownian) particle 
\begin{equation}
    \langle x^2\rangle=2 D t
\label{eq:D}
\end{equation} 
may be generalized naturally to the 
many-body setting with use of correlation/green function\cite{Chaikin_Lubensky_1995}. Briefly, we compute the probability distribution of a disturbance in the particle density field at time $t$ and compute its second spatial moment accordingly. Working with the  conventional dynamic structure factor, Eq. \ref{eq:D} becomes
\begin{equation}
        \lim_{Q\to 0}\partial^2 S(Q,t)/\partial Q^2= D t.
\end{equation}
This clearly illustrates the importance of the so-called "transport" limit of taking $Q$ (and $1/L$) to zero first at finite time. Note, that while one often observes $S(Q,t)\propto \exp(-D Q^2 t)$, this is not required for the existence of diffusion -- the standard definition is only concerned with the rate of the initial decay of $S(Q,t)$.   The complementary regime of late times at finite $Q$ allows for interactions among diffusive modes to manifest\cite{raj2024diffusion}. There is a crossover time $t_Q\to \infty$ as $Q\to 0$ that separates the short-time diffusive behavior from late time interaction dominated regime. 

To be sure, there are important perturbative effects of nonlinearities already in the transport regime: the conventional LTTs \cite{subrotoLTT,dorfmanLTT,ernst1984long} usually refer to conductivity and (via Einstein relation) in the temporal corrections to the diffusion constant itself $D\to D(t)=D_\infty + A/t^\gamma+\ldots$.  Here the non-conserved current should relax fast (ala Drude formula) but instead acquires power-law tail(s) through perturbative mixing with slow variables -- the effect is usually easily captured in low order perturbation theory even when it produces quantitatively large corrections\cite{subrotoLTT,dorfmanLTT,lucadima24} (but not always\cite{oganesyan2009energy}).  Such perturbative mechanism is absent (at least in leading order) in models without disorder and with only one conserved charge as noted in Ref.\cite{subrotoLTT} -- these conditions are satisfied in the model we study here.

Instead, we find\cite{raj2024diffusion,heavyluca} a superficially similar but wholly non-perturbative phenomenon of the diffusion cascade which allows for excitations at with finite $Q$ to redistribute the momentum into several excitations.  These collective eigenstates of $H$ form a continuum that can reach arbitrarily slower decays than elementary single mode excitations at any finite $Q$, provided $1/L\to \infty$.
%and by doing so live longer, usually powerlaw LTT but only stretched exponential in this model. 
To illustrate this mechanism, suppose a single particle mode with momentum $Q$ fragments into 4 modes, each with momentum $Q/4$, thus reducing its decay rate from $D Q^2$ to $4 D (Q/4)^2= DQ^2/4$ and so on (in an infinite system). For the model considered, these processes occur at progressively higher orders in perturbation theory (in nonlinearity)\cite{heavyluca} and therefore take some time to build up, e.g. if we initialize the dynamics in a single Fourier mode with definite $Q$ \cite{raj2024diffusion}.  Thus, as long as the correct order of limits is taken ($Q\to 0$ first), these processes effectively never take place and can be ignored. 

Now, returning to the main goal of this paper -- analytic calculation of the diffusion constant -- we recognize the challenge for exact (or approximate) spectral methods -- once we focus on eigenvalues, we, by definition, obtain asymptotic late time decay of finite $Q$ states, which is ideal if we are interested in studying the diffusion cascade (see Fig. 5 in Ref.\cite{raj2024diffusion}) but requires additional care for obtaining the diffusion constant directly.  The only ground state that correctly represents the diffusion constant is the lowest allowed momentum sector $Q_1=2\pi/L$ (for periodic boundary conditions) as single mode excitations with that momentum have nowhere to decay  -- no cascade. To be sure, \emph{real-time dynamics} initialized in all other sectors generically displays correct diffusion decays at short times but these eventually cross over  to the slower decay dominated by interactions $D Q_n^2\to D Q_n^2/n$(see Fig. 5 of Ref.\cite{raj2024diffusion}).
%This can be used to compute for example the density-density correlation functions using $O=n_1 n_{x+1}$. We see that correlation functions will decay exponentially at a rate which is the inverse of the lowest energy on which $P(0)$ has overlap with. For example, if the initial configuration $P(0)$ is built with using only eigenfunctions with momentum $Q$, the decay time will be the inverse of the ground state momentum in the sector $Q$, $E_{1,N}=E_{0,N,Q}$
%\begin{equation}
%t_{eq}(Q)=1/E_{0,N,Q}.
%\end{equation}
%We expect $E_{0,N,Q}\simeq D(\rho) Q^2+O(Q^4)$ for small $Q$ and we observe $D(\rho)$ that decreases monotonically with $\rho$. The global ground state $E_{0,N}$ is in the $Q=0$ sector, obviously. The Hamiltonian is real and positive, the ground state will correspond to zero decay rate.  Importantly, the existence of the diffusion cascade\cite{raj2024diffusion} restricts straightforward extraction of the diffusion constant in a finite system to the lowest momentum $Q=2\pi/L$, which is what we use henceforth.
Fortunately, in what follows, we will be able to blissfully ignore this subtlety and estimate the diffusion constant simply by computing the dispersion (mass) of the effective single quasiparticle state using mean-field theory.  The cascade processes are associated with the existence of many-body continua which do not exist in the simplest mean-field theory we employ.

\section{Fermionization and a mean field approximation}

In order to proceed with the analysis of the model, we break $h_{i,j,k}$ into Pauli basis\footnote{These first technical steps are similar to \cite{schulz1997analytical}} and first rewrite the rate matrix as the Hamiltonian of a spin system with $s^\alpha=\sigma^\alpha/2$ where $\sigma^\alpha$ are Pauli matrices:
\begin{eqnarray}
    h_{1,2,3}&=&\left(\frac{1}{2}-s^z_1\right)\vec{s}_{2}\cdot\vec{s}_{3}+\left(\frac{1}{2}-s^z_{2}\right)\vec{s}_{3}\cdot\vec{s}_{1}+\nonumber\\
    &+&\left(\frac{1}{2}-s^z_{3}\right)\vec{s}_{1}\cdot\vec{s}_{2}
    +\frac{1}{3}\left(s^z_1+s^z_2+s^z_3-\frac{3}{2}\right).
\end{eqnarray}
This form is also reminiscent of a PXP spin model on a triangular ladder\footnote{Recognizing that 
$\left(\frac{1}{2}-s^z_i\right)\vec{s}_{i+1}\cdot\vec{s}_{i+2}=\left(\frac{1}{2}-s^z_i\right)\vec{s}_{i+1}\cdot\vec{s}_{i+2}\left(\frac{1}{2}-s^z_i\right)$,
and similarly for the other terms. Explicitly, $H=-\sum_i P_iX_{i+1,i+2,i}P_i +P_{i+1}X_{i+2,i,i+1}P_{i+1}+P_{i+2}X_{i,i+1,i+2}P_{i+2}$,
with $P_i=\left(1/2-s^z_i\right)$ and $X_{a,b,c}=\vec{s_a}\cdot\vec{s_b}+P_c$.}. Then we re-write it in terms of hard-core bosons, by defining: $f=\sigma^-,f^\dag=\sigma^+$, $\sigma^z=2f^\dag f-1$. After some algebra we have:
\begin{eqnarray}
\label{eq:A_f}
    h_{1,2,3}&=&\frac{1}{2}\left(f_1 f_2f_2^\dag f_3^\dag+f_1^\dag f_2 f_2^\dag f_3+f_1f_2^\dag f_3 f_3^\dag+f_1^\dag f_2 f_3 f_3^\dag+f_1 f_1^\dag f_2 f_3^\dag\right.\nonumber\\
    &+&\left.f_1 f_1^\dag f_2^\dag f_3\right)
    -f_2 f_2^\dag f_3 f_3^\dag-f_1 f_1^\dag f_3 f_3^\dag-f_1 f_1^\dag f_2 f_2^\dag+\nonumber\\
    &+3&f_1 f_1^\dag f_2 f_2^\dag f_3 f_3^\dag.
\end{eqnarray}
As usual, hard-core bosons $f_i$ in one dimension can be fermionized via a Jordan-Wigner transformation
$$\sigma^+_j = f_{j}^{\dag }=e^{-i\pi \sum _{k=1}^{j-1}a_{k}^{\dag }a_{k}}\cdot a_{j}^{\dag } 
 =\prod _{k=1}^{j-1}(1-2a_k^\dag a_k)\cdot a_{j}^{\dag }$$
$$ \sigma^-_j = f_{j}=e^{+i\pi \sum _{k=1}^{j-1}a_{k}^{\dagger }a_{k}}\cdot a_{j}=\prod _{k=1}^{j-1}(1-2a_k^\dag a_k)\cdot a_{j}$$
$$ f_{j}^{\dagger }f_{j}=a_{j}^{\dagger }a_{j} = \frac{\sigma^z_j + I}{2}$$
to get  
\begin{eqnarray}
\label{eq:A_a}
    h_{1,2,3}&=&\frac{1}{2}\left(a_1^\dag a_2a_2^\dag a_3 + a_3^\dag a_2a_2^\dag a_1 + a_1^\dag a_3 a_3^\dag a_2 + a_2^\dag a_3 a_3^\dag a_1 \right.\nonumber\\
    &+&\left. a_2^\dag a_1a_1^\dag a_3 + a_3^\dag a_1a_1^\dag a_2 \right) \nonumber\\
    &-&a_2 a_2^\dag a_3 a_3^\dag-a_1 a_1^\dag a_3 a_3^\dag-a_1 a_1^\dag a_2 a_2^\dag
    \nonumber\\
    &+3&a_1 a_1^\dag a_2 a_2^\dag a_3 a_3^\dag.
\end{eqnarray}
The full Hamiltonian is obtained  by summing these terms over all the triangular plaquettes. The resulting fermionic Hamiltonian then can be split into a ``conditioned" hopping on the triangle in Fig.\ref{fig:hopping}:
\begin{equation}
    T=\frac{1}{2}\sum_{ \langle i,j,m \rangle } a^\dag_i(1-a^\dag_j a_j)a_m+\mathrm{h.c.},
\end{equation}
where the notation $\langle i,j, m\rangle$ denotes three sites that belong to a triangle, by analogy with standard nearest-neighbor notation $\langle i, j \rangle$.
There is also a mixed 2-3 fermion density interaction term
\begin{equation}
    V=-\sum_{
    \langle i,j\rangle}(1-a^\dag_i a_i)(1-a^\dag_j a_j)+3\sum_{\langle i,j,m\rangle}(1-a^\dag_i a_i)(1-a^\dag_j a_j)(1-a^\dag_m a_m),
\end{equation}
and finally
\begin{equation}
    H=T+V.
    \label{eq:Ham_Fermi}
\end{equation}

Constrained hopping quantum particles are another interesting subject of recent research in particular for their slow but not many-body localized dynamics, see for example\cite{sierant2021constraint}. This Hamiltonian, written in this form, does not seem to be exactly solvable despite being clearly a special case (it is {\it stoquastic}, i.e.\ the off-diagonal terms are all of the same sign, and does not have a sign problem either). So, in order to make progress we must make some approximations. Let us see how much mileage we get from a simple mean field approximation. To start, let us consider the ground state. 
In a given sector $N$, the exact ground state is the superposition of all possible configurations of $N$ particles in $L$ sites. We have
\begin{equation}
    \ket{E_{0,N}}=\frac{1}{Z}\sum_{1\leq j_1\leq j_2\leq ... j_N\leq L}a^\dag_{j_1}a^\dag_{j_2}...a^\dag_{j_N}\ket{0}.
\end{equation}
If we define the {\it grand-canonical state} the coherent state
\begin{equation}
    \ket{\Psi(z)}=e^{z a^\dag_1}...e^{z a^\dag_L}\ket{0},
\end{equation}
then it is not difficult to see that
\begin{equation}
    \ket{E_{0,N}}=\frac{1}{Z}\oint\frac{dz}{2\pi i}z^{-N-1}\ket{\Psi(z)}.
\end{equation}
In the large $N,L$ limit, one can use the saddle point approximation\footnote{This means one can use the saddle point approximation for any observable, which is equivalent to using the approximated vector.} and obtain
\begin{equation}
    \ket{E_{0,N}}\simeq\frac{1}{Z^{1/2}}\ket{\Psi(z_s)},
\end{equation}
where $z$ is the {\it fugacity} at the saddle point (and $Z=(1+|z|^2)^L$ is a normalization constant) and it is fixed by the density $\rho=N/L$:
\begin{equation}
    \bra{E_{0,N}}\sum_{x=1,...,L}a^\dag_x a_x\ket{E_{0,N}}=N,
\end{equation}
and
\begin{equation}
    \bra{\Psi(z)}\sum_{x=1,...,L}a^\dag_x a_x\ket{\Psi(z)}=L\frac{|z|^2}{1+|z|^2}=N.
\end{equation}
So $|z|^2=\rho/(1-\rho)$ with an arbitrariness of the phase $\theta=\arg(z)$.

The grand-canonical state $\ket{\Psi(z)}$ is then a good state for starting a mean field approximation, with fixed $\rho=N/L$ (instead of fixed $N$, but as usual the fluctuations will not matter) which is good not only at small $\rho$ but, as it turns out, also in whole range of $\rho$. 

First of all, consider that the state is factorized so the expectation values of product of operators on different $x,y$ sites will become the product of expectation values. Therefore it is sufficient to see that
\begin{eqnarray}
    \langle a^\dag_x\rangle_{{\rm MF}}&=&z^*/(1+|z|^2),\nonumber\\
    \langle a_x\rangle_{{\rm MF}}&=&z/(1+|z|^2),\nonumber\\
    \langle a^\dag_x a_x\rangle_{{\rm MF}}&=&\rho,\nonumber\\
    \langle a^\dag_x a_y\rangle_{{\rm MF}}&=&\rho(1-\rho)\ \mathrm{ for }\ y\neq x.
\end{eqnarray}
Inserting these into Eq.(\ref{eq:Ham_Fermi}), we find
\begin{equation}
\bra{\Psi(z)}H\ket{\Psi(z)}=\frac{1}{2}(1-\rho)6\rho(1-\rho)-3(1-\rho)^2+3(1-\rho)^3=0.
\end{equation}
As we know that the exact GS energy is $E_{N,0}=0$ in any sector $N$ we now understand what it means. $\Psi(z)$ is a weighted superposition of all the GS's in all the sectors. By relaxing the particle number conservation we can proceed with the calculation of the spectrum of the quaisparticle of the model. 

The MF treatment of excitations on top of the ground state can be obtained in the Hartee-Fock approximation (see for example \cite{wen2004quantum}, Chapter 5) by replacing the local density terms $a^\dag_i a_i$ terms with their expectation value on the MF GS, we therefore obtain the effective Hartree-Fock Hamiltonian
\begin{equation}
    H_{\mathrm{HF}}=\sum_{\langle i,j,m\rangle}\frac{1}{2}(1-\rho)\left(a_i^\dag a_j + a_j^\dag a_i + a_i^\dag a_m + a_m^\dag a_i + a_j^\dag a_m + a_m^\dag a_j \right)-3(1-\rho)^2+3(1-\rho)^3,
\end{equation}
where $i,j,m$ refer to real space lattice positions. The last term is the interaction energy estimated in MF, a constant
while the first term is the hopping on a triangular ladder, which we has dispersion law that is easiest to write using Fourier space coordinates ($k$) of a chain twice the length with nearest and next-nearest neighbor hopping (hence the presence of both $cos(k)$ and $\cos(2 k)$ terms):
\begin{equation}
    \epsilon_k=(3-\cos(2k)-2\cos(k))(1-\rho)\simeq 3(1-\rho) k^2.
\end{equation}
The quasi-particles therefore have an effective mass $m=\frac{1}{6}(1-\rho)^{-1}$ which diverges as $\rho\to 1$. The quasiparticles are created by acting with $a^\dag_k$ on the coherent state $\ket{\Psi(z)}$ and they have energy $\epsilon_k=\bra{\Psi(z)}a_k H a^\dag_k\ket{\Psi(z)}.$ 
The HF Hamiltonian for the quasiparticle excitations $a_k\equiv L^{-1/2}\sum_j e^{i k j} a_j$ is
\begin{equation}
    H_{\mathrm{HF}}=\sum_{k}\epsilon_k a^\dag_k a_k-3(1-\rho)^2+3(1-\rho)^3.
    \label{eq:HF_Ham}
\end{equation}
Therefore $[a^\dag_k,H]=\epsilon_k a_k^\dag$ which shows that $\epsilon_k$ determines the decay rate of the state $a^\dag_k\ket{\Psi(z)}$.

We can now go back to the random dynamics of the classical particles by extracting the diffusion coefficient from the lowest energy $\epsilon_{Q}$, to the first excited state which is in the $Q=2\pi/L\ll 1$ sector as $\epsilon_{Q}=D_1 Q^2+...$. The MF prediction for the diffusion coefficient is then:
\begin{equation}
    D_{{\rm MF}}=3(1-\rho).
    \label{eq:DMF}
\end{equation}
This simple form matches the single particle result (see $D_0$ in Section 2.1) and appears to capture very well the numerical data in Sec. \ref{sec:numx}, up to large density $\rho\sim 5/6$. However, for $\rho>1/2$ and in particular $\rho\gtrsim 2/3$ the system size dependence becomes important, presumably due the reduction of the number of accessible configurations.
%decreasing considerably. 
In fact, we will see that an exponential number of configurations becomes exactly immobile or {\it jammed} for $\rho \geq 2/3$ (see Sec. \ref{sec:jammedstats}, next) which is manifest in finite time crossover to true diffusion but also in finite size dependence in spectral data, as we document in Sec. \ref{sec:numx}.

\section{Large densities and the appearance of jammed configurations}
\label{sec:jammedstats}
As we increase the density past the critical value $\rho=2/3$ a series of particle configurations appear which are stuck or jammed, namely that cannot be moved by our dynamical rules. They are trivial ground states of the Hamiltonian since $H\ket{\sigma}=0$ for each of these. At exactly $\rho=2/3$ one can see that the single configuration $\sigma=110110110...$ and its 2 translates by $T$ are jammed. At this point 3 more ground states appear in the spectrum. \footnote{Notice that the total number of such configurations of length $L$, {\it not constrained} to have a given density $\rho$, is a known combinatorial problem which has as solution the sequence A000930 of OEIS, also known as Narayana's cows sequence: 1, 2, 3, 4, 6, 9, 13, 19, 28, 41, 60, 88,...\ . This number grows asymptotically like $x^L$ where $x=1.4655...$ is the only real solution of the equation $x^3-x^2+1=0$. Notice also that this is exactly the exponential of the maximum of the entropy Eq.(\ref{eq:s_lock}), $x=\exp(s_{j}(\rho^*))$ which is achieved at $\rho^*=0.80574$. So the typical frozen configurations have density $\rho=\rho^*$ and there are $x^L$ of them.}

As $\rho>2/3$ the number of such configurations becomes exponential in $L$, and therefore we define an entropy of jammed configurations $s_{j}=(\ln \mathcal{N})/L$. Computing the entropy of such configurations $\mathcal{N}$ using the tools of statistical mechanics is not difficult: Fixing the first vertex to be empty (therefore quotienting wrt to translations which introduces at most a factor $L$ which does not change the entropy), we have sequence of $n_1\geq 2$ ones, then an isolated 0, then a sequence of $n_2\geq 2$ ones, then another zero and so on. So the number of sequences of 1 is equal to the number of zeros $N_0=L-N$.

\begin{figure}
    \centering
    \includegraphics[width=0.5\linewidth]{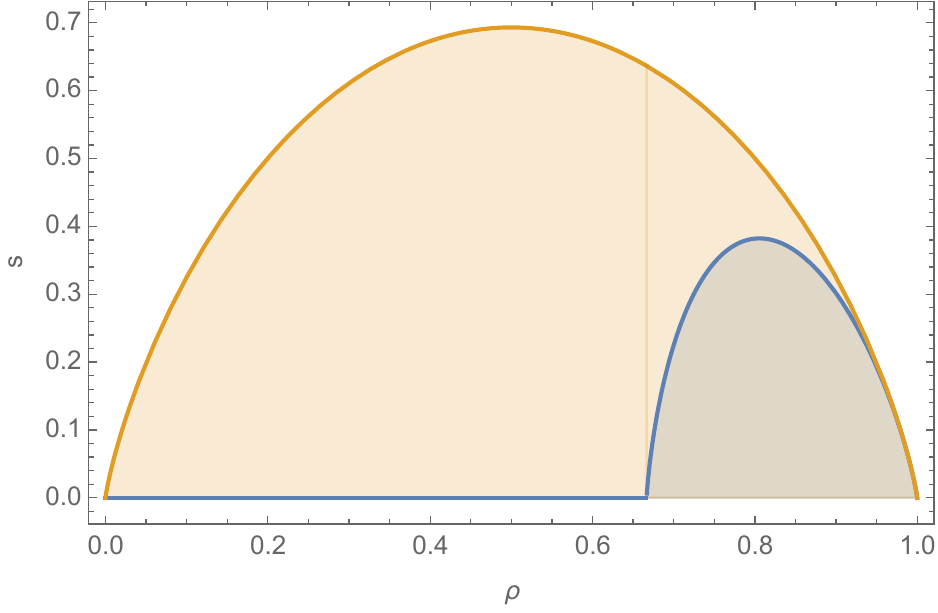}
    \caption{Entropy of jammed configurations (blue) and total entropy of the system (yellow).}
    \label{fig:entropy_jam}
\end{figure}

So we can compute the number of sequences of $N_0=L-N$ numbers $n_i\geq 2$ that sum to $N$. So defining the set of integer sequences $\mathbb{S}=\{\{n_i\}_{i=1}^{N_0}| n_i\geq 2 \}$
\begin{equation}
    \mathcal{N}(N)=\sum_{\{n_i\}\in\mathbb{S}}\delta_{\sum_{i=1}^{N_0} n_i,N}.
\end{equation}
We need to compute this number for large $N,L$ fixing $\rho=N/L$. To do this we first compute the generating function (grand-canonical ensemble)
\begin{equation}
    \mathcal{N}(z)=\sum_{N=0}^\infty \mathcal{N}(N)z^N,
\end{equation}
which we will eventually invert using:
\begin{equation}
    \mathcal{N}(N)=\oint \frac{dz}{2\pi i}z^{-N-1}\mathcal{N}(z).
\end{equation}
Now, $\mathcal{N}(z)$ is an unconstrained geometric series:
\begin{equation}
    \mathcal{N}(z)=\sum_{n_1\geq 2, ...,n_{N_0}\geq 2} z^{\sum_{i=1}^{N_0}n_i}=\frac{z^{2N_0}}{(1-z)^{N_0}}.
\end{equation}
So 
\begin{equation}
    \mathcal{N}(N)=\oint \frac{dz}{2\pi i}z^{-N-1}\frac{z^{2N_0}}{(1-z)^{N_0}}
\end{equation}
and we can evaluate this integral for large $N_0=(1-\rho)L$ using the saddle point method. After some straightforward calculations one gets
\begin{equation}
    \mathcal{N}(N)\simeq e^{Ls(\rho)}
\end{equation}
with
\begin{equation}
\label{eq:s_lock}
    s_{j}(\rho)=(\rho -1) \log \left(\frac{1-\rho }{2 \rho -1}\right)+(2-3 \rho ) \log \left(\frac{3 \rho-2 }{2 \rho -1}\right).
\end{equation}
This curve is depicted in Fig.\ref{fig:entropy_jam}, together with the total entropy $s(\rho)=-\rho\ln\rho-(1-\rho)\ln(1-\rho)$. It reaches a maximum at a value $\rho^*=0.80574...$.
As the density is increased, one can see that a larger fraction of the configurations become jammed, as the difference between the total entropy and the entropy of the jammed configurations goes to zero (see Figure \ref{fig:entropy_jam}) as $s(\rho)-s_{j}(\rho)\simeq 2(\rho-1)^2+...$. 

If the initial probability distribution $P(0)$ has overlap with one of these, then it will {\it never} fully relax to the equipartite equilibrium $P_{eq}=\frac{1}{Z}\sum_\sigma\Ket{\sigma}=\frac{1}{Z^{1/2}}\ket{E_0}$, but rather leave behind a "localized fraction"\cite{localizedfraction}, a jammed fingerprint of the initial state, whose weight will be exponentially small $\log  \mathcal{N}/Z=2(1-\rho)^2+3(1-\rho)^2+\ldots$. %\textcolor{red}
%{[AS: OK for me to remove it!]}\VO{***Is it ok to remove this? it seems redundant/obvious but also sloppy, i.e. we need to again specify lowest Q and probably another caveat.****
%\\
%However, if the initial configuration {\it does not} have overlap with any jammed configuration, it will decay to the equilibrium state in a time which is again dictated by diffusion $T_{eq}=L^2/D$. The value of $D$ must be extracted from numerics. It turns out, again, to be well described by the MF result $D=3(1-\rho)$, see Fig.\ref{fig:Particle_sector}, although with relevant finite-size corrections.}

\section{Dynamics near maximum density: holes and doublons (and more-ons)}
\label{sec:doublons}
We now turn to exploiting spectral methods to compute the dynamics of few hole excitations of the fully occupied inert state of $N=L$ particles ($N_h\equiv L-N$).  To outline the strategy: (i) recall, that at low hole density $\epsilon=1-\rho$ the leading contribution to thermodynamics and dynamics is from the single hole sector, $N_h=1$ (to be precise, random arrangements of far separated, non-interacting holes).  
Since the model is constructed to render these immobile (and absent long-range interaction or hopping processes) we are guaranteed that the diffusion constant vanishes $D(\epsilon)\to 0$ as $\epsilon\to 0$; (ii) To go further we need to consider the influence of other excitations on these holes. The leading contender to generate "slow holes" is the "doublon" which we loosely identify as configurations with two "nearby" holes that evade kinetic constraints and diffuse, we calculate the exact doublon spectrum below, including its diffusion constant $D_2=3/8$; (iii) Any configuration that contains doublons (or generally not restricted to single holes) is dynamical. The lowest energy (i.e. longest living) states are ones containing just one doublon, with $N_h-2$ slow holes. One may argue for its rate of relaxation to vary $\sim (N_h-2) D_2 /L^3$ by either appealing to the real space picture of diffusive doublon  spreading out a single hole over typical distance between holes $~L/(N_h-2)$ or by working out the dispersion many-body states (i.e. the continuum) at Fourier momentum $Q=2\pi/L$. We defer a rigorous treatment of these interactions effects to future work (see also numerical results in Sec. \ref{sec:numx} for $N_h=3,4,5$). If valid at finite hole density, these results imply $D(\epsilon)\sim \epsilon$.

%The density of doublons is $\epsilon^2$, which implies $D\sim \epsilon$. 

%Near $\rho=1$ we can still find a diffusion coefficient. 
%\subsection{Frozen holes}
%We consider the sectors $N=L,L-1,L-2,...$ . The sector $N=L$ has only one state, the completely full one: this state is clearly jammed. The sector $N=L-1$ has $N$ states (the positions of the single hole) but they are all jammed as well since one needs two holes at least for them to be mobile.

\subsection{Mobile doublons}
\begin{figure}
    \centering
    \includegraphics[width=0.5\linewidth]{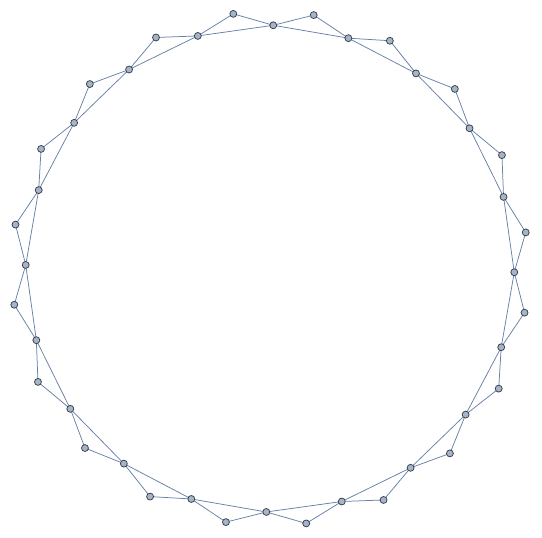}
    \caption{Reduced Hamiltonian of the $L=20,N=18$ configuration. It contains $2L=40$ unjammed configurations, which are connected either to 2 or 4 other configurations. Amplitudes are $-1/2$ over all the off-diagonal matrix elements.}
    \label{fig:DoublonH}
\end{figure}

The sector $N=L-2$ shows the first unjammed configurations, although the holes have to move together to be mobile, separated by at most one particle.  So we have $2L$ unjammed configurations: $L$ for separation $0$ and $L$ for separation $1$. We call these configurations a {\it doublon}.  The configuration with two holes separated by a particle $...1101011...$ can move in two possible ways: $...1110011...,\ ...1100111...$, while the configuration with two neighboring holes $...110011...$ can move in 4 possible ways. All amplitudes are $-1/2$ and the diagonal terms are $1$ and $2$ respectively (to ensure $\sum_i H_{ij}=0$). The graph of possible motions is as in Figure \ref{fig:DoublonH}.

The effective Hamiltonian of the doublon is now a chain decorated with triangles. Numbering the vertices $i=1,...,2L$, the equations for the doublon amplitudes $\psi_i$ are
\begin{eqnarray}
    E\psi_{i+2}&=&2\psi_{i+2}-\frac{1}{2}\psi_{i+1}-\frac{1}{2}\psi_{i}-\frac{1}{2}\psi_{i+3}-\frac{1}{2}\psi_{i+4}\\
    E\psi_{i+1}&=&\psi_{i+1}-\frac{1}{2}\psi_{i}-\frac{1}{2}\psi_{i+2}.
\end{eqnarray}
Separating the odd (connected to 2) and even (connected to 4) configurations, and using the plane wave ansatz $\psi_{j,o}=a_1 e^{ijk}, \psi_{j,e}=a_2 e^{ijk}$ respectively, with now $j=1,...,L$ we find the eigenvalues by diagonalizing the reduced matrix
\begin{equation}
    M=\left(
\begin{array}{cc}
 1 & -\frac{1}{2}-\frac{e^{i k}}{2} \\
 -\frac{1}{2}-\frac{e^{-i k}}{2} & 2-\cos (k) \\
\end{array}
\right).
\end{equation}
The eigenvalues form a sound and an optical branch (in the first the particles move together, in the second they oscillate around a common center of mass):
\begin{equation}
    \epsilon_{k}=\frac{1}{4} \left(-2 \cos (k)\pm\sqrt{2} \sqrt{\cos (2 k)+7}+6\right).
    \label{eq:doublon_sp}
\end{equation}
At small $k$ we have
\begin{eqnarray}
    \epsilon_{k,1}&=&\frac{3}{8}k^2+O(k^4),\\
    \epsilon_{k,2}&=&2+\frac{1}{8}k^2+O(k^4).
\end{eqnarray}
Remembering that the minumum $k=Q=2\pi/L$ we find that the diffusion coefficient of 2 holes is
\begin{equation}
    D_{N=L-2,L}\equiv\epsilon_{k,1}\frac{L^2}{(2\pi)^2}\to\frac{3}{8}.
\end{equation}
So the diffusion coefficient is small but not zero for $N=L-2$. This compares perfectly with the numerics (blue points in Fig.\ref{fig:high_density}).

\begin{figure}
    \centering
    \includegraphics[width=0.5\linewidth]{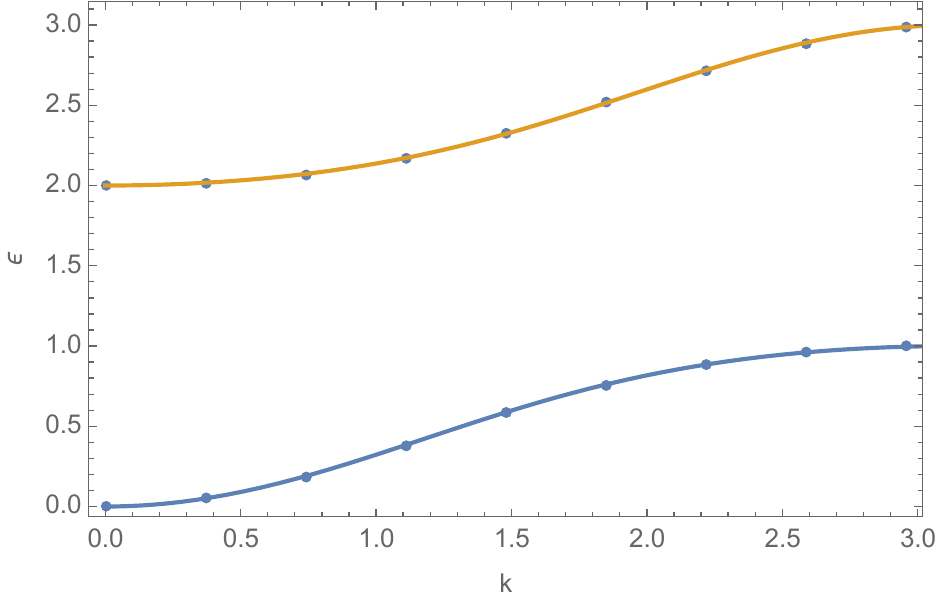}
    \caption{Energy vs momentum of the doublons eigenstates, in the case $N=15,L=17$. Numerics (dots), and analytics (lines) in Eq.(\ref{eq:doublon_sp}).}
    \label{fig:enter-label}
\end{figure}

\section{Exact numerical results}
\label{sec:numx}
The classical model can be simulated by a random process in which on picks a random triangular plaquette and then, if it qualifies, i.e.\ it contains a single particle, makes a random move of the particle on one of the other two, empty, vertices (if it does not qualify, then nothing happens). Such a random process, quick to implement, leads to correlation functions decay among observables\cite{raj2024diffusion}. Although, the decays are eventually exponential, as governed by gaps discussed above, this is not a good way to extract gaps due to typically dismal signal-to-noise ratio\cite{raj2024diffusion}. 
%\footnote{As already mentioned in Sec. \ref{sec:model} the existence of the diffusion cascade obstructs naive $D(\rho)  Q^2$ dispersion of the decay rate. Briefly, while initial temporal decay may follow $D(\rho) Q^2 $ the late time decay rate is much slower, $\sim Q^2/n$, where $n=L Q/2\pi\geq 1$. This is the focus of a companion paper\cite{raj2024diffusion}}

%the longest time-scale being the equilibrium time $T_{eq}$ discussed above. 

Alternatively, one can write a (sparse) rate matrix, and find the gap above the ground state with Lanczos-like algorithms. In this way we can access mesoscopic ($L=30$) to large ($L=200$) system sizes, depending on the number of particles. We have preferred this latter method, by which we have produced the figures in this section.

\subsection{Low density limit}

\begin{figure}
    \centering
\includegraphics[width=0.49\columnwidth]{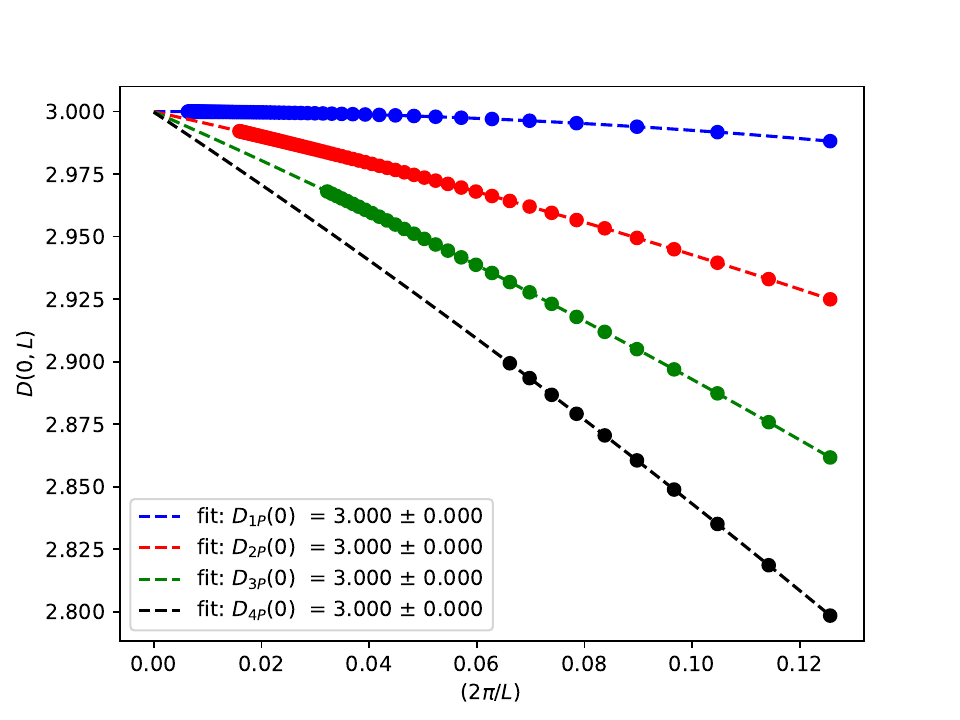}
    \caption{Diffusion coefficient vs $Q=\frac{2\pi}{L}$ for $\rho \rightarrow 0$ with quadratic fit. $L=50, 60,\cdots 990$ for 1 particle (blue). $L =  50,55, \cdots 395$ for 2 particles(red). $L = 50, 55, \cdots 195$ for 3 particles(green). And $L = 50,55, \cdots 95$ for 4 particles(black). They all extrapolate to the theoretical value $D=3$ valid for density $\rho=0$.}
    \label{fig:low_density}
\end{figure}

In the low-density limit we expect the prediction of the fermionic MF theory to be correct. $D=3(1-\rho)$, in particular $D=3$ for $\rho=0$, which is reproduced by $O(1)$ particles as $L\to\infty$ as can be seen in \ref{fig:low_density}. Notice the different number of particles give rise to different $1/L$ corrections but not the final result.

\begin{figure}
    \centering
\includegraphics[width=0.49\columnwidth]{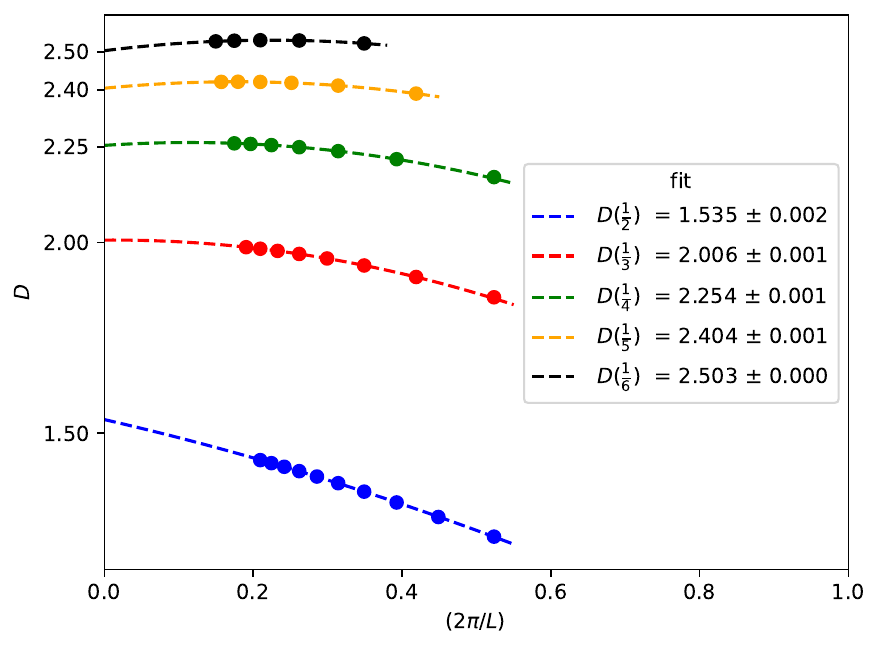}
\includegraphics[width=0.49\columnwidth]{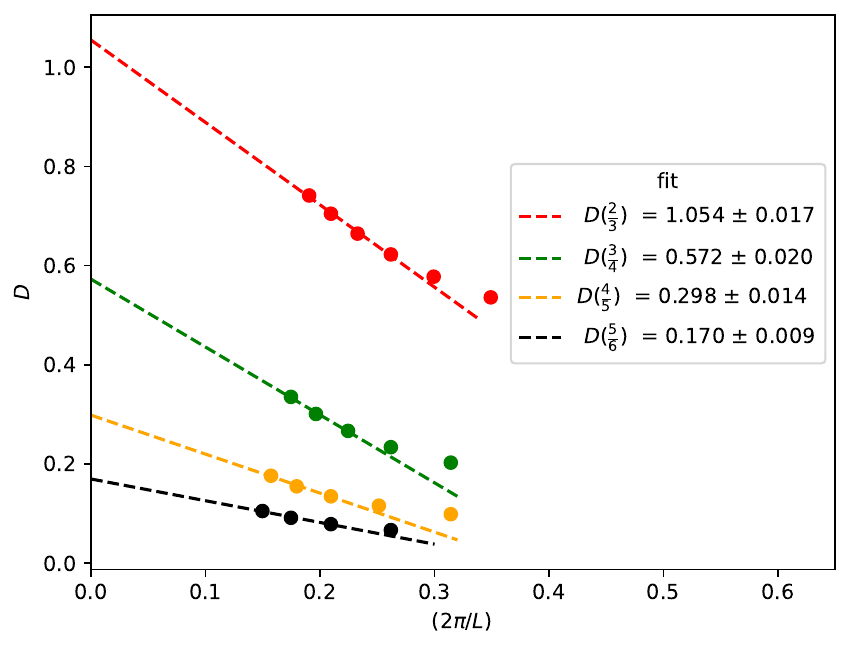}

    \caption{{\it (Left)} Diffusion coefficient vs $Q=\frac{2\pi}{L}$ with a quadratic fit and the values predicted by MFT $D=3(1-\rho)$. $L=8, 10,\cdots 30$ for $\rho = 1/2$ (blue). $L=9, 12,\cdots 33$ for $\rho = 1/3$ (red). $L =  12,16, \cdots 36$ for $\rho =1/4$ (green). $L = 15, 20, \cdots 40$ for $\rho =1/5$(orange). And $L = 18,24, \cdots 42$ for $\rho =1/6$(black). {\it (Right)} Diffusion coefficient vs $Q=\frac{2\pi}{L}$ with linear fit. $L=18, 21,\cdots 33$ for $\rho = 2/3$ (red). $L =  20,24, \cdots 36$ for $\rho =3/4$ (green). $L = 20,25, \cdots 40$ for $\rho =4/5$(orange). And $L = 24, 30, \cdots 42$ for $\rho =5/6$(black).}
    \label{fig:low_density_1bn}
\end{figure}

Going to larger densities the result does not change substantially, and the prediction $D=3(1-\rho)$ is satisfied up to $\rho=1/2$ as seen in Figure \ref{fig:low_density_1bn}, although again, larger densities correspond to large finite-size corrections.

\subsection{After jamming and towards isolated doublons}
\label{sec:numx2345}
\begin{figure}
    \centering
\includegraphics[width=0.49\columnwidth]{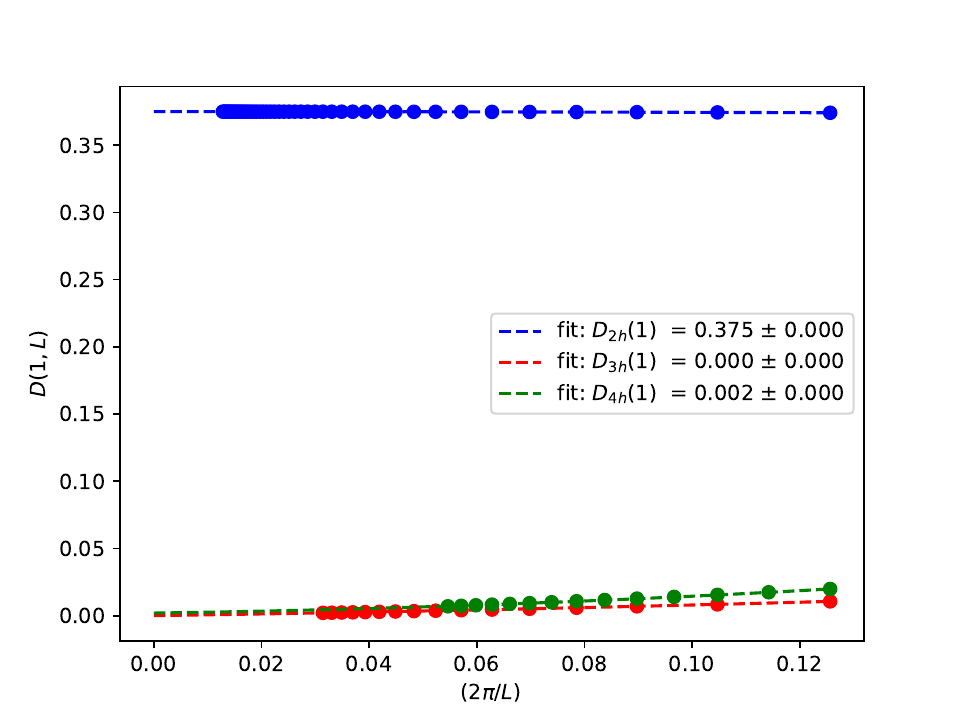}
\includegraphics[width=0.49\columnwidth]{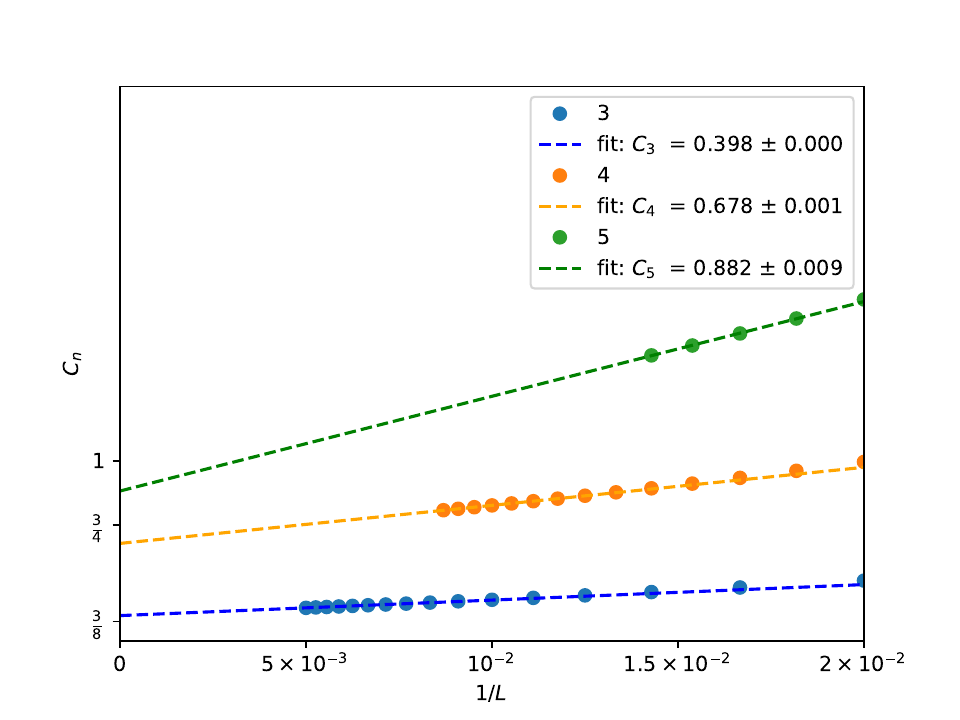}
    \caption{{\it (Left)} Diffusion coefficient vs $Q=\frac{2\pi}{L}$ for $\rho \rightarrow 1$ with quadratic fit. $L=50, 60,\cdots 490$ for 2 holes (blue). $L =  50, 60, \cdots 200$ for 3 holes(red). $L = 50, 55, \cdots 115$ for 4 holes(green). {\it(Right)} For $n\geq 3$ holes the coefficient $c_n$ defined ad $c_n=LD_{N-n,L}$ vs $1/L$ and the extrapolated valued.}
    \label{fig:high_density}
\end{figure}

Crossing the transition at $\rho=2/3$, where one starts to have jammed configurations, the finite-size corrections become important. This might be due to the fact that the number of accessible configurations from an initial, unjammed one, can be considerably smaller than expected, which typically increases the finite-size correction. A simple linear extrapolation in $1/L$ does agree with the prediction of MF theory within an acceptable $5\%$ margin. See figs.\ \ref{fig:low_density_1bn}. However, the limit $N=L-n$ with $L\to\infty, n=O(1)$ falls back into the analysis of the previous section. The diffusion coefficient for $n=2$ is $D=3/8$ while for $n\geq 3$ is $D_n=c_n/L$ with $c_3=3/8, c_4=3/4$ and we conjecture $c_n=(3/8)[n/2]$ where $[n/2]$ is the integer part of $n/2$, which counts the number of doublons. However, already for $n=5$ the finite-size corrections are so large that we cannot confirm this prediction (see Fig.\ref{fig:high_density}, right panel).

\section{Conclusions and further work}
\label{sec:fin}
We have designed and solved, in the mean-field approximation, a kinetically constrained model of particles hopping on a triangular ladder. The solution is obtained by a classical-quantum mapping to a model of interacting fermions. The diffusion coefficient is the inverse of the mass of the quasiparticles, which can be computed in the mean-field approximation and returns a monotonically decreasing diffusion coefficient in striking agreement with the numerics in a large interval of densities. Two sets of numerics have confirmed these results: both Montecarlo calculations on the classical particle model and exact diagonalization of the quantum Hamiltonian. As directions for further work, one should try to either go beyond the mean-field approximation or, since we see little deviations from the MF result in the numerics, prove its exactness. Moreover, the model can be easily generalized to a triangular lattice in 2 dimensions \footnote{We thank Frank Pollman for this suggestion, which opens the way to an unexpected generalization of our work.} where one still finds jammed configurations and the quantum model is a model of hard-core bosons with conditioned hopping. We also notice that a classical-quantum mapping on a fermionic model, like the one devised in this paper, might be useful to get a field-theory approach to other KCPs or facilitated spin models, which is fundamentally different from other known ones, which are based on mode-coupling theory (see \cite{pitts2000facilitated,perrupato2022exact} and references therein) or other mappings to many-body models, followed by a different kind of approximation \cite{schulz1997analytical}.

\section{Acknowledgements}
A.S.\ would like to thank the Graduate Center of CUNY for hospitality during a visit in July 2024, when this project got started. The work of A.S.\ was funded by the European Union - NextGenerationEU under the project NRRP “National Centre for HPC, Big Data and Quantum Computing (HPC)CN00000013 (CUP D43C22001240001) [MUR Decree n.\ 341- 15/03/2022] - Cascade Call launched by SPOKE 10 POLIMI: “CQEB” project.  A.R. and V.O. would  like to thank Sarang Gopalakrishnan and Paolo Glorioso for collaboration on a related work\cite{raj2024diffusion}. The authors would like to thank P.~Krapivski for comments on the first version of this paper. 

\section{Bibliography}
\bibliography{references}
\bibliographystyle{plain}
\end{document}